\documentclass[journal]{IEEEtran}

\usepackage{cite}
\usepackage{amsmath,amssymb}
\usepackage{graphicx}
\usepackage{textcomp}
\usepackage{xcolor}
\usepackage{url}
\usepackage{hyperref}

\def\j23{\textsc{J23}}
\def\Z12{\textsc{Z12}}
\def\HD24{\textsc{HD25}}

\newcommand{\UCLA}{Department of Physics and Astronomy, University of California, Los Angeles, CA 90025, USA}
\newcommand{\CIT}{California Institute of Technology, 1200 E. California Blvd., Pasadena, CA 91125, USA}
\newcommand{\JPL}{Jet Propulsion Laboratory, California Institute of Technology, Pasadena, CA 91109, USA}
\newcommand{\ASU}{School of Earth and Space Exploration, Arizona State University, Tempe, AZ 85287, USA}

\newcommand{\UPENN}{Department of Physics and Astronomy, University of Pennsylvania, Philadelphia, PA 19104, USA}
\newcommand{\UIUCPHYS}{Department of Physics, University of Illinois Urbana-Champaign, Urbana, IL 61801, USA}
\newcommand{\UVA}{Department of Astronomy, University of Virginia, Charlottesville, VA 22904, USA}
\newcommand{\PRINCETON}{Department of Physics, Princeton University, Princeton, NJ 08544, USA}
\newcommand{\NCSA}{NCSA, University of Illinois Urbana-Champaign, Urbana, IL 61801, USA}
\newcommand{\UIUCASTRO}{Department of Astronomy, University of Illinois Urbana-Champaign, Urbana, IL 61801, USA}
\newcommand{\ASUSEMTE}{School for Engineering of Matter, Transport and Energy, Arizona State University, Tempe, AZ 85287, USA}
\newcommand{\UCHICAGOASTRO}{Department of Astronomy and Astrophysics, University of Chicago, Chicago, IL 60637, USA}
\newcommand{\CFA}{Center for Astrophysics, Harvard \& Smithsonian, Cambridge, MA 02067, USA}
\newcommand{\MPIA}{Max-Planck-Institut fur Astronomie, D-69117 Heidelberg, Germany}
\newcommand{\UARIZONA}{Department of Astronomy, University of Arizona, Tucson, AZ 85721, USA}
\newcommand{\UCHICAGOPHYS}{Department of Physics, University of Chicago, Chicago, IL 60637, USA}
\newcommand{\UIUCJOINT}{Department of Astronomy, Department of Physics, NCSA, University of Illinois Urbana-Champaign, Urbana, IL 61801, USA}
\newcommand{\KICP}{Kavli Institute for Cosmological Physics, University of Chicago, 5640 South Ellis Avenue, Chicago, IL 60637, USA}
\newcommand{\FNAL}{Fermi National Accelerator Laboratory, MS209, P.O. Box 500, Batavia, IL 60510, USA}
\newcommand{\UmassLowell}{Department of Physics and Applied Physics, University of Massachusetts, Lowell, Lowell, MA 01854}

\title{Kilopixel Performance of the Kinetic Inductance Detectors for the Terahertz Intensity Mapper}

\author{Justin S. Bracks,
Reinier M. J. Janssen,
Steve Hailey-Dunsheath,
Talia Saeid,
Bruce Bumble,
Logan Foote,
Elijah Kane,
Lun-Jun Liu,
Charles M. Bradford,
Shubh Agrawal,
James E. Aguirre,
Hrushi Athreya,
Brockton S. Brendal,
Karia Dibert,
Peter Dow,
Jeffrey P. Filippini,
Jianyang Fu,
Karolina G. Garcia,
Christopher E. Groppi,
Bradley Johnson,
Dylan Joralmon,
Wooseok Kang,
Garrett K. Keating,
Ryan P. Keenan,
Ian N. Lowe,
Alex Manduca,
Aashrita Mangu,
Daniel P. Marrone,
Philip D. Mauskopf,
Evan C. Mayer,
Sydnee O'Donnell,
Simon Tartakovsky,
Mathilde Van Cuyck,
Joaquin D. Vieira,
and Jessica A. Zebrowski

\thanks{J. S. Bracks is with \UCLA{} and \CIT{}.}%
\thanks{R. M. J. Janssen, L. Foote, E. Kane, and C. M. Bradford are with \JPL{} and \CIT{}.}%
\thanks{B. Bumble is with \JPL{}.}%
\thanks{S. Hailey-Dunsheath, L.-J. Liu, and K. Dibert are with \CIT{}.}%
\thanks{T. Saeid is with \ASU{} and \JPL{}.}%
\thanks{S. Agrawal, J. E. Aguirre, and A. Manduca are with \UPENN{}.}%
\thanks{H. Athreya, B. S. Brendal, J. P. Filippini, and S. O'Donnell are with \UIUCPHYS{}.}%
\thanks{P. Dow and B. Johnson are with \UVA{}.}%
\thanks{J. Fu and S. Tartakovsky are with \PRINCETON{}.}%
\thanks{K. G. Garcia is with \NCSA{}.}%
\thanks{C .E. Groppi is with \UmassLowell{}.}%
\thanks{P. D. Mauskopf is with \ASU{}.}%
\thanks{D. Joralmon is with \ASUSEMTE{}.}%
\thanks{W. Kang is with \UCHICAGOASTRO{}.}%
\thanks{G. K. Keating is with \CFA{}.}%
\thanks{R. P. Keenan is with \MPIA{}.}%
\thanks{I. N. Lowe, D. P. Marrone, and E. C. Mayer are with \UARIZONA{}.}%
\thanks{A. Mangu is with \UCHICAGOPHYS{}.}%
\thanks{M. Van Cuyck is with \UIUCASTRO{}.}%
\thanks{J. D. Vieira is with \UIUCJOINT{}.}%
\thanks{J. A. Zebrowski is with \KICP{} and \FNAL{}.}%
}

\begin{document}

\maketitle

\begin{abstract}
We characterize a flight-grade array of lumped-element kinetic inductance detectors (LEKIDs) developed for the long-wavelength module of the Terahertz Intensity Mapper (TIM). From an 864-pixel science array, we select 490 well-isolated resonators spanning the focal plane and readout band and measure their thermal response, optical responsivity, and noise using a  ZCU111-based multitone readout system intended for flight. The thermal and optical response of the detector population is consistent with previous single-pixel measurements and can be described by Mattis-Bardeen theory under the influence of a change in Cooper pair or quasi-particle density. Under increasing blackbody loading, the measured noise transitions from a  thermal generation-recombination-dominated floor to photon-noise-limited scaling, with the majority of detectors achieving photon-noise limited operation by approximately 400 fW incident power, with a median detector noise limited NEP of $NEP_{det} = 1.1\times10^{-17} \ \mathrm{W \sqrt{Hz}}$. The measured photon-noise scaling implies a median optical efficiency of approximately 0.67, indicating additional unknown loss sources between the detectors and blackbody radiator, tentatively attributed to losses in the waveguide. These results demonstrate that the TIM LEKID architecture retains the required sensitivity when scaled to kilopixel-class arrays; the principal remaining challenges are array-level tone optimization, resonator tracking, and identification of frequency-domain collisions.

\end{abstract}


\section{Introduction}
\IEEEPARstart{T}{he} key diagnostic deliverable of a line intensity mapping (LIM) experiment is not a high resolution image showcasing the visible characteristics of individual astronomical bodies. Instead, LIM surveys generate a power spectrum of the image. The LIM power spectrum probes correlations in the intensity field as a function of spatial scale. This generates a measurement of the aggregate emission from unresolved source populations and characterizes the resulting three-dimensional intensity field statistically. This allows astronomers to probe the large-scale structure of the universe and gather statistics on vast populations of galaxies. \cite{changLidz26} give an excellent review.

The Terahertz Intensity Mapper (TIM) is a balloon-borne LIM telescope that targets a specific emission line from singly ionized Carbon, [CII] at 158 $\mathrm{\mu m}$ rest-frame wavelength. The Doppler shift imposed on this cosmic light imprints time-of-flight information into the observed wavelength itself, giving TIM the ability to generate 3D maps of the ancient universe. We motivate and present a detailed accounting of TIM's scientific capabilities in \cite{bracks26}.

While TIM has no need to resolve individual galaxies, the angular resolution of an intensity map does set the lower bound of the spatial scales that can be probed in the angular directions. Similarly, the spectral resolution of the spectrometer determines the smallest line of sight (radial) scale available. Taken together, this defines a minimum probeable cosmic volume, a voxel. Increasing the number of spatial and spectral pixels decreases the statistical error on a measurement. Similarly, an intensity mapper must accurately measure the in-band photon power within each voxel. Each of these requirements pressure the need for more densely populated, high-sensitivity far-infrared detector arrays. TIM will field eight roughly kilopixel arrays of lumped element kinetic inductance detectors (LEKIDs). This work extends the single-pixel characterization of Janssen et al. \cite{janssen23} (\j23 hereafter) to a flight-grade kilopixel array and tests whether the previously demonstrated detector performance is preserved when the detector count is increased by approximately an order of magnitude.

In this project we characterize a statistically representative population of LEKIDs in a kilopixel array designed for use in TIM's long wavelength module. In Section \ref{sec:experiment} we document our experimental setup. In Section \ref{sec:thermal_measurements} we discuss the un-illuminated behavior of the array as a function of cryogenic temperature. Then, in Section \ref{sec:optical_measurements} we explore the behavior of the LEKIDs while illuminated with a flight-like optical loading and use the optical measurements to help characterize the noise performance of our KIDs compared to scientific specifications. In Section \ref{sec:discussion} we discuss the detector properties and performance. Finally, in Section \ref{sec:conclusion} we summarize and conclude.


\section{Experimental Setup}
\label{sec:experiment}
\begin{figure}[!t]
    \centering
    \includegraphics[width=\columnwidth]{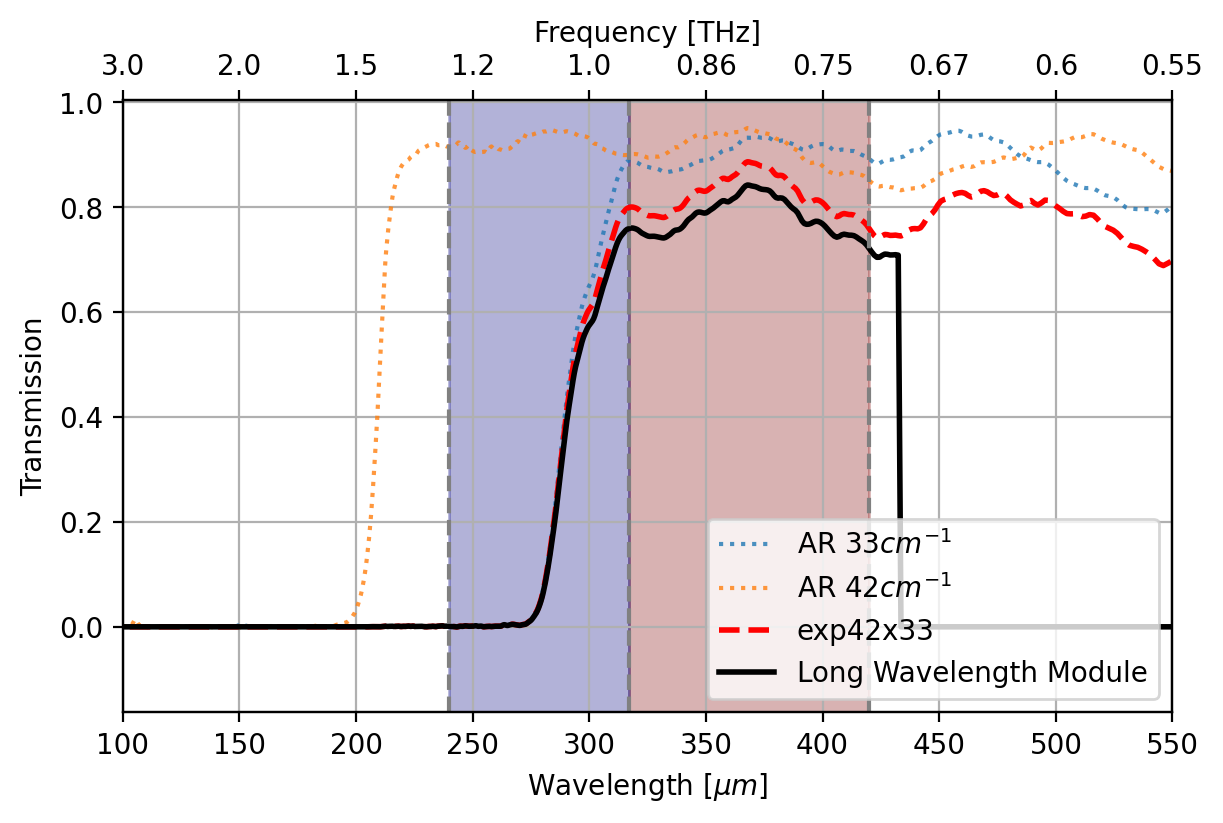}
    \caption{Transmission as a function of wavelength across TIM’s science band. Orange and blue dotted lines are the transmission curves of the two low-pass filters in our testing optics. Red dashed is their product. Black line is the total transmission curve for our experimental setup, including the waveguide cut-off. Blue and red fill regions depict TIM’s short- and long-wavelength bands, respectively.}
    \label{fig:filter}
\end{figure}
\subsection{Array Design and Fabrication}
We measure the performance of a TIM flight-grade array for the long wavelength spectrometer arm. This array consists of 864 all aluminum LEKIDs and readout line patterned by a single front-side lithographic step into a 30 nm film. Individual detectors closely match the design presented in \j23, consisting of an inductor following the ``chain-link'' design and an interdigitated capacitor. The chain-link absorber \cite{Nie2022} consist of a 500 nm wide meandering line, which can achieve a $\sim 90\%$ band averaged absorption efficiency when combined with optical choke rings and a quarter-wavelength backshort. Following \j23, the backshort is produced using deep-trench reactive ion etching (DRIE) of the silicon-on-insulator (SOI) substrate. The DRIE is used to etch a 1.3 mm diameter hole through the handle wafer after which the entire backside is metallized using 200 nm aluminum. This creates a backshort 25 $\mathrm{\mu m}$ below the absorber as defined by the SOI device layer thickness. 

The aluminum interdigitated capacitors have 2 $\mathrm{\mu m}$ wide tines spaced by 4 $\mathrm{\mu m}$ gaps. Each detector has a unique resonance frequency set by adjusting the number of tines and tine lengths. The design of these tine properties follows a unit cell approach \cite{Liu2024,Foote2024}, in which 16 KIDs (4 by 4 block) each have a unique tine count, maximally spreading them in the target readout bandwidth (500 - 1000 MHz). Each of the 54 instances of the unit cell in the array has unique tine lengths, thus creating 864 unique resonators.

A single feedline meanders through the array \cite{Liu2022} to which all KIDs are capacitively coupled. The most significant change with respect to the J23 design, is the increase of this coupling capacitance to reduce the coupling Q, $Q_c$, to $5\times10^4$. This reduction was done to better match the internal quality factor, $Q_i$, expected in flight based on updated instrument and flight-loading models \cite{FuPhDthesis}. 


\subsection{Cryogenics and Optics}
Our cryogenic system consists of a pulse-tube pre-cooled triple-stage He sorption cooler. The pulse-tube provides a 50 K intermediate stage and a 4 K environment. The sorption cooler enables the experimental stage to achieve a $\sim 210$ mK base temperature, despite the 4K radiative environment.
The array is mounted in a single array version of the housing design presented in Liu et al. \cite{Liu2022} and placed on the detector stage. Held between 50 $\mathrm{\mu m}$ bosses and spring-loaded pogo pins, the wafer is affixed beneath a precision-machined, aluminum horn block. Each pixel is fed by a single two-flare angle horn and 254 $\mathrm{\mu m}$ diameter, 481 $\mathrm{\mu m}$ long waveguide. The precision-machined feedhorns are tightly arranged in a hexagonal pattern to maximize active area. Mounted directly on top the hornblock are metal-mesh low-pass filters whose product gives a $>70\%$ transmission across TIM's LW band. The first of the filters is a 1 THz $(\lambda \approx 300 \mu m)$ cut-off low-pass filter. The second is a 1.3 THz $\lambda \approx 220 \mu m$ low-pass. Together with the cut-off frequency of the circular waveguide at 700 GHz, these filters create a band-pass around the TIM long-wavelength spectrometer science band. Figure \ref{fig:filter} shows the individual contributions to the filter transfer functions (dashed \& dotted lines) and the combined transmission as a function of frequency (solid black line).

Above the horn block a large cryogenic blackbody made from Tessellating TeraHertz RAM tiles \cite{THzTiles} on a copper backing plate is affixed on the inside of the 4K shell using weak thermal links. Through a series of three resistive heaters, the blackbody can be heated to $>10$ K without significant increase in the 4K environmental temperature, thus providing the variable in-band loading for the optical loading measurements presented in this work.

\subsection{Multiplexed Microwave Readout}

\begin{figure*}[!t]
\centering\includegraphics[width=\textwidth]
{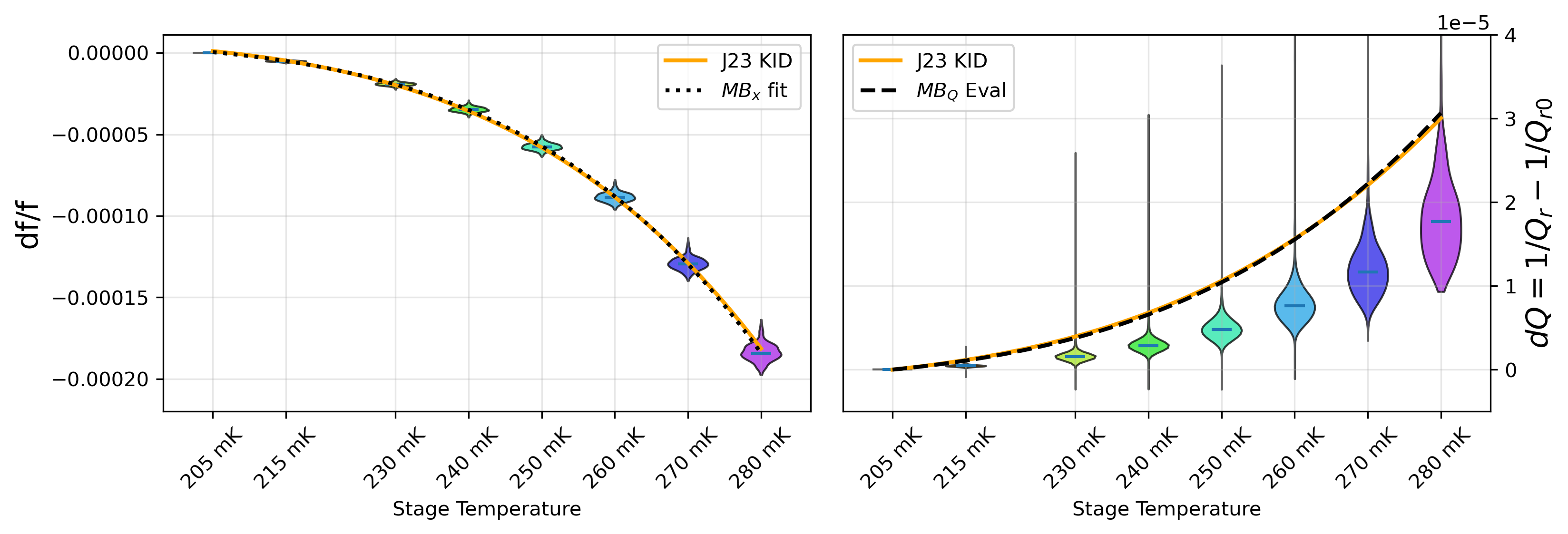}
        \caption{On the left: fractional frequency shift due to increasing cryogenic bath temperature. On the right: change in quality factor. Blue horizontal lines report the median values of the data. The orange solid lines are the best-fit curves from \j23 where $\alpha_{\rm kin}$ was held at 0.68. The black dotted lines in the left panel is the best-fit curve resulting from fitting to the median df/f data to \ref{e: x} and allowing both $\alpha_{\rm kin}$ and $\Delta_0$ to be fit freely. The dashed line in the right panel is an evaluation of Eq \ref{e: Q} using the best-fit parameters obtained by that fit.}
    \label{fig:thermal}

\end{figure*}

Detector readout is performed using an FPGA readout system based on the Xilinx ZCU-111 radio frequency system on a chip (RFSoC). The RFSoC is capable of generating 1,000 tones in a 500 MHz bandwidth for each of 4 readout circuits – sufficient to readout the entire focal plane. However, the dynamic range of the RFSoC digital to analogue converter is not sufficient to stimulate the resonators across the dynamic loading range of our variable optical power experiment. 
For our optical measurements, we insert a simple variable amplification chain just after the RFSoC analogue output. This consists of a LabBrick variable attenuator ahead of a ZKL-1R5+ 20 dB amplifier (For the dark measurements described in the following section, we remove this attenuation and control drive power only using the RFSoC output). This allows us to balance the input power on the amplifier to avoid compression and control the total power at the cryostat input. Inside the cryostat, the signal lines are thermally anchored at 50K, 4K, 300 mK and the experimental stage. A 30dB and 10 dB attenuator are present at 4K and 300mK, respectively, on the input line to minimize thermal noise. A low noise $(T_{noise} = 4 \ \mathrm{K})$ cryogenic amplifier is located on the 4K stage and provides 30 dB gain before the signal exits the cryostat. Following the cryostat output, we include a static 3 dB attenuator ahead of a VLF3400+ low pass filter to avoid amplifying out-of-band signal and minimize gain ripples and standing waves between the amplifiers. Finally, the signal passes through a chain of 2 LNAs (a Miteq LNA30-00100400-13-10P  and then another ZKL-1R5+), each followed by a LabBrick variable attenuator. This allows us to condition the power at the RFSoC analogue to digital converter over the dynamic signal power range needed to conduct our optical measurements. 

\section{Dark Measurements}
\label{sec:thermal_measurements}
Initial characterization of the array using a dedicated Vector Network Analyzer (VNA) as well as similar capabilities on the RFSoC-based readout, we identify 776 individual resonances features, suggesting a fabrication yield of 759/864 KIDs (88\%). The resonators' unilluminated resonant frequencies at nominal flight-time stage temperature of 250 mK, span 481.4 MHz to 982.6 MHz with a single outlier at 1004 MHz. We measure a loading-agnostic median $Qc = 10^{(4.7\pm 0.2)}$. These values closely match the design and are within TIM's operation requirements.
 
To test our kilopixel array in detail we first identified a statistically representative population of the KIDs that: 
\begin{itemize}
    \item were individually well isolated in terms of their resonance frequency, 
    \item constituted full spatial coverage across the surface of the array, 
    \item constituted full frequency coverage across TIM's LW band.
\end{itemize}  

Conservatively defining `well isolated' as having no near (resonant frequency) neighbor within 100 kHz yielded a population of 490 KIDs that fulfilled the other bullets above. The above bullets were intended to ensure that the population of KIDs were both truly representative of the array and (relatively) easy to characterize across a range of optical loadings and cryogenic temperatures.

For these dark measurements the blackbody heaters are unbiased. In essence, the detectors are staring at 4K environment, which we actually measure to be at 3.5 K and provides a limited photon loading as supported by results from our optical measurements Sect~\ref{sec:noise_performance}. Figure \ref{fig:nep_vs_power} illustrates that at 3.5 K loading ($\sim6\times10^{-13}$ W) detector noise power is largely dominated by generation-recombination with a $\sim 9.5\%$ contribution from photon shot noise. There is an additional $\sim 1\%$ contribution from amplifier-injected noise which we determine via the ratio between amplitude and phase noise.

We determine each resonator's individual quality factor, $Q_{res}$ and resonant frequency, $f_{res}$, as a function of the detector stage temperature.
Figure \ref{fig:thermal}  shows the measured change in resonance frequency (left frame), $x(T) = (f_{res}(T) - f_{res}(0)) / f_{res}(0)$, and quality factor (right frame) $1/Q_{res}(T) - 1/Q_{res}(0)$ as a function of stage temperature, $T_S$, for 490 well-isolated KIDs. As in \j23, we model the thermal response of our KIDs following the standard application of Mattis-Bardeen theory \cite{zmuidzinas12}:

\begin{equation}
x_{MB} = \frac{\alpha_{kin}\gamma S_2}{4N_0 \Delta_0}n_{qp}
\label{e: x}
\end{equation}

\begin{equation}
Q_{MB}^{-1} = \frac{\alpha_{kin}\gamma S_1}{2N_0 \Delta_0}n_{qp}
\label{e: Q}
\end{equation}

We maintain uniformity with \j23 in all notation and parameter values in equations \ref{e: x} and \ref{e: Q}. From \j23:
\begin{quote}
Where $\alpha_{kin}$ is the kinetic inductance fraction and $\gamma = 1$ is appropriate for the thin films used here. $S_1$ and $S_2$ are the standard expressions as given in Eq. 71 and 72 of \Z12. The quasi-particle density, $n_{qp}$ is given by $n_{th} = 2N_0 \sqrt{2\pi k_bT\Delta_0}$  $exp[ -\Delta_0 / k_bT]$.
We adopt a density of states $N_0 = 1.72\times10^{10} \ \mathrm{\mu m^{-3} \ eV^{-1}}$ \cite{gao08} and the BCS relation between the gap energy, $\Delta_0 = 1.76$ $k_bT_c$, and transition temperature, $T_c$.
\end{quote}

We see strong agreement in the evolution of $f_{\rm res}$ with regard to the measurements presented in Figure 2 of \j23 with a marginally reduced resonator response to stage temperature. However, we note a significant dampening in the evolution of $dQ$ compared to \j23. We tentatively attribute this to a difference in microwave readout power, which has known non-equilibrium interactions with the quasi-particles at high densities \cite{deVisser2014b}. \j23 kept the stimulus constant over increasing temperature, whilst here we optimize the readout to remain close to bifurcation. Furthermore, the reduction in $Q_c$ allows for an increased coupling of stimulus power into the device. Following these thermal response measurements, we fit the median $df/f(T)$ data to equation \ref{e: x}. Allowing the fitting algorithm to vary both $\alpha_{\rm kin}$ and $\Delta_0$ we retrieve (expected) $\Delta_0$--$\alpha_{\rm kin}$ degenerate solutions. However, we know from the work in \j23 that $\alpha_{\rm kin} \approx 0.68$ and  $\Delta_0 \approx 205$ $\mu$EV. Indeed, holding $\alpha_{\rm kin}$ fixed at this predetermined value returns the corresponding predetermined $\Delta_0$ with an excellent fit to the median data values. We then evaluate equation \ref{e: Q} with the fit parameters obtained from the frequency response and find the dashed black curve in the right-hand panel of Figure \ref{fig:thermal}.


\section{Optical Measurements}
\label{sec:optical_measurements}

\subsection{Responsivity}

\begin{figure*}[!t]
\centering\includegraphics[width=\textwidth]
{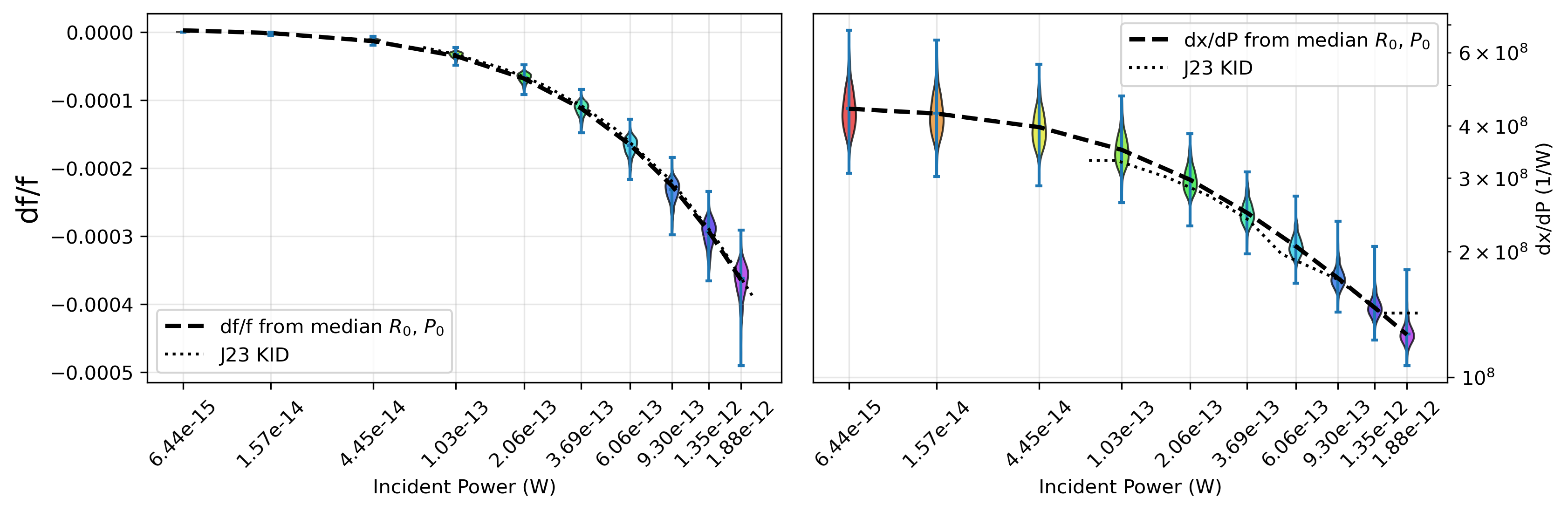}
        \caption{$df/f_0$ (left) and responsivity (dx/dP) (right) as functions of $P_{\rm inc}$. Dashed black lines are the per resonator median $R_0$, and $P_0$ fit values determined by fitting each KID's data to \ref{eq: fFrac} (left) and \ref{eq: dxdP} (right). The dotted black line reproduces the single-pixel measurement \j23.} 
    \label{fig:response}

\end{figure*}

To characterize the optical response and noise of the KIDs we regulate the stage temperature at 250 mK, then illuminate the array using the cryogenic blackbody discussed in section \ref{sec:experiment}. The blackbody reaches a minimum temperature of $T_{BB} \approx 3.5\ \mathrm{K}$ and was increased by 0.5 K steps up to 7.5 K. The power incident on the absorber can be calculated as:
\begin{equation}
    \int N_m n_{BE}(T_{BB}) F(\nu) h \nu d\nu
\end{equation}

Where $\nu$ is frequency; $h$ is Planck's constant; $F(\nu)$ is the frequency dependent filter transfer function discussed above; $n_{BE}$ is a Bose-Einstein distribution that defines the photon occupation number and $N_m$ describes the total number of modes. 

From here we begin to follow the prescription for optical measurements laid out in Hailey-Dunsheath et al. \cite{shd24} (\HD24 hereafter) Sect 4B. For the 250 mK stage temperature and optical loading considered in these experiments \cite{zmuidzinas12,deVisser2014b}, the quasi-particle generation due to absorption of microwave readout power is negligible, hence, the fractional frequency shift of a resonator's resonant frequency as a function of incident power, $x = f_{res}(P_{inc}) - f_{res}(0) / f_{res}(0)$, may be written as \cite{zmuidzinas12,shd24}:

\begin{equation}
    dx/dP_{inc} = R_0[1 + \frac{P_{inc}}{P_0}]^\frac{1}{2}
    \label{eq: dxdP}
\end{equation}

using:
\begin{equation}
    R_0 = \eta_{opt} \frac{\alpha_{kin} \gamma S_2(\omega)}{4 N_0 \Delta_0} \frac{\eta_{pb} \tau_{max}}{\Delta_0 V}(1 + \frac{n_{th}}{n^*})^{-1}
\end{equation}
and
\begin{equation}
    P_0 = \frac{n^* \Delta_0 V}{2 \eta_{pb} \eta_{opt} \tau_{max}}(1 + \frac{n_{th}}{n*})^2.
\end{equation}

Where $\eta_{opt}$ is the absorber optical efficiency describes the fraction of incident power that is actually absorbed in the detector and $\eta_{pb}$ is the pair-breaking efficiency describes the fraction of photon energy that goes into Cooper pair breaking. V is the inductor volume. $\tau_{max}$ and $n^*$ are constants determined by the material properties such that $\tau_{qp} = \tau_{max}(1+ n_{qp}/n^*)^{-1}$, where $\tau_{qp}$ is the quasiparticle lifetime and $n_{qp}$ is the quasiparticle number density for the optical load in question. Integrating Eq. \ref{eq: dxdP} and assuming that both $P_0$ and $R_0$ are constant with $P_{inc}$ (which is true for $P_0$ and effectively true for $R_0$ across our loading regime) yields:

\begin{equation}
    \frac{df}{f} = 2 R_0 P_0 [(1 + \frac{P_{inc}}{P_0})^\frac{1}{2} - 1]
    \label{eq: fFrac}
\end{equation}

Having measured $df/f (P_{inc}$) for all KIDs in our representative population, we first fit $R_0$ and $P_0$ for each individual resonator and, applying Eq. \ref{eq: dxdP}, calculate the detectors' responsivity, $dx/dP$. 
Figure \ref{fig:response} reports the fractional frequency shift, $x = df/f$, (left frame) and responsivity $dx/dP(P_{inc})$ of our KIDs in response to optical loading. We compare the population-level statistics with a representative KID from our \j23 paper. The population is in good agreement with the single-pixel prototype presented in \j23. In the fractional frequency shift plot we see a nearly one-to-one overplot with the population median, and in the dx/dP plot the \j23 KID is within the population scatter. Here we also demonstrate that when the individual resonators are well tracked across the loading regime the measurement spread is significantly reduced.  

\subsection{Noise performance}
\label{sec:noise_performance}

We measure the fractional-frequency noise, $S_{xx}$, of each resonator by recording on-resonance timestreams at fixed microwave tone power. The complex transmission timestreams are first rotated into the frequency-response direction using the local resonator sweep. We then calculate the power spectral density of the fractional frequency fluctuation,
\begin{equation}
    x(t) \equiv \frac{f_{\rm res}(t)-\langle f_{\rm res}\rangle}
    {\langle f_{\rm res}\rangle},
\end{equation}
for each optical loading. We quote the white-noise amplitude from the approximately flat portion of the spectrum surrounding 10 Hz, chosen to avoid low-frequency excess noise from the readout chain. At its 488 Hz sampling rate, the ZCU-111 based multi-tone readout system is not fast enough to capture high-frequency roll-off in the noise spectrum.

\begin{figure}[h]
\centering\includegraphics[width=\columnwidth]
{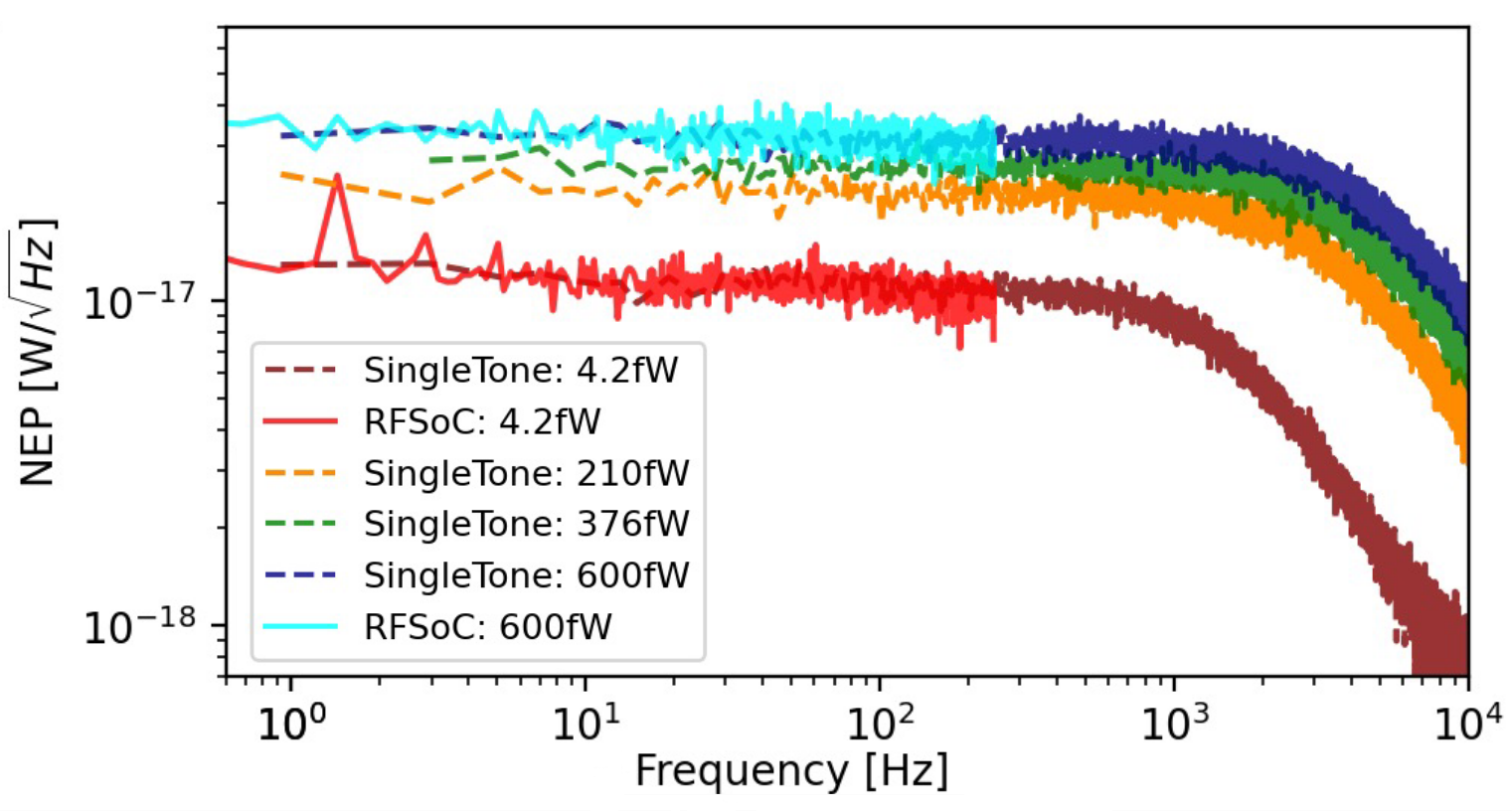}
        \caption{Example noise power spectral densities of one representative KID. Data were obtained both with a single-tone homodyne readout (see \j23), which is fast enough to capture the resonator response time (roll-off), as well as the RFSoC multitone readout used in this study, and which will be used during TIM's science flight. Spectra of both the detector noise (dark conditions, reds) and under maximum expected TIM loading (blues) are presented along with two intermediate loading examples.} 
    \label{fig:noise_psd}
\end{figure}

Figure~\ref{fig:noise_psd} shows representative fractional-frequency noise spectra for several blackbody temperatures. We provide high sampling rate measurements using the single-tone homodyne readout system described by \j23 for comparison. At the lowest optical loading the spectra are consistent with thermal generation-recombination (GR) noise over the science band. As the blackbody temperature is increased, the white-noise level rises while preserving an approximately flat spectrum over the band of interest. This behavior is consistent with the transition to Poisson noise from photons. The absence of a strong frequency dependence in the science band indicates that two-level-system noise, which typically has a $1/\sqrt(f)$ spectrum \cite{gao08}, or $1/f$ noise from the readout chain, does not substantively contribute to the measured detector noise at 250 mK.

Combining the measured fractional-frequency noise with the optical responsivity from Sect.~\ref{sec:optical_measurements}A, the noise equivalent power referenced to incident optical power is
\begin{equation}
    {\rm NEP}_{\rm inc}(f)
    =
    \frac{\sqrt{S_{xx}(f)}}{\left|dx/dP_{\rm inc}\right|}
    \sqrt{1+\left(2\pi f \tau_{\rm qp}\right)^2}.
    \label{eq:nep_inc}
\end{equation}
The first two factors in Eq.~\ref{eq:nep_inc} are directly measured: $S_{xx}$ from the timestreams and $dx/dP_{\rm inc}$ from the optical response fits. The final term accounts for the finite quasiparticle lifetime, $\tau_{\rm qp}$, which suppresses the response at high modulation frequency. In the present work the relevant science band is well below the inverse quasiparticle lifetime, so
the lifetime correction is negligible for the quoted low-frequency NEPs.  

For comparison with the measured NEP, we calculate the photon-limited noise expected for the optical power incident on the detector. Following the convention
used in \HD24, the photon-limited NEP referenced to absorbed optical power is
\begin{equation}
    {\rm NEP}_{\gamma}^{2}
    =
    2 h \nu P_{\rm abs}
    \left(
        1+n_0+\frac{2\Delta_0}{h\nu\eta_{\rm pb}}
    \right) +K,
    \label{eq:photon_nep_abs}
\end{equation}
where $P_{\rm abs}=\eta_{\rm opt}P_{\rm inc}$, $n_0$ is the photon occupation number in the detector mode and $K$ the detector noise contribution. The three terms in parentheses
correspond to photon shot noise, photon wave noise, and quasiparticle recombination noise, respectively. In the TIM long-wavelength band the photon
occupation number is small for the blackbody temperatures considered here, and as such the wave-noise term is subdominant. We therefore expect ${\rm NEP}_{\rm inc}\propto P_{\rm inc}^{1/2}$ when photon and recombination
noise dominate.


The principal goal of this measurement is to determine whether a large, flight-representative population of TIM long-wavelength KIDs can be driven and
read out simultaneously while maintaining the sensitivity required for the TIM science case. For each detector in the 490-KID analysis population, we compute the optical responsivity using Eq.~\ref{eq: dxdP} and convert the measured white-noise level to ${\rm NEP}_{\rm inc}$ using Eq.~\ref{eq:nep_inc}. 

\begin{figure*}[!t]
    \centering
    \includegraphics[width=0.95\textwidth]{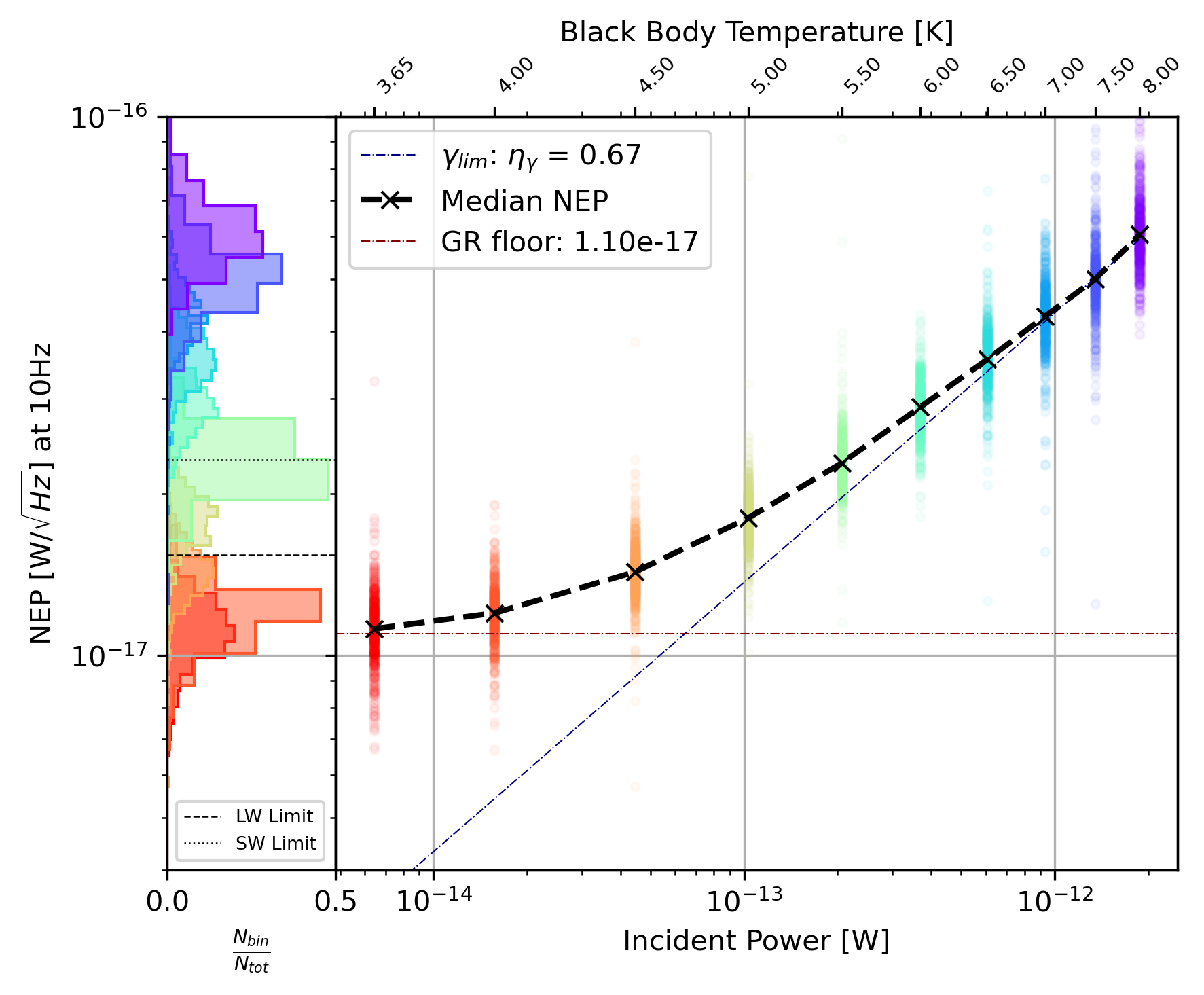}
    \caption{NEP($P_{\rm inc}$) for 490 selected TIM KIDs in one flight-grade long-wavelength array depicted as histograms per loading regime (left) and scatter (right). The black X's represent the median value at each loading. Horizontal, maroon dashed-dotted line is the thermal noise floor, and the diagonal navy dash-dot line is the fit to the photon shot noise term. In the left panel, the dotted black line is the SW science spec, and the dashed black line (sitting on top of the maroon dash-dot line) is the LW science spec.
    }
    \label{fig:nep_vs_power}
\end{figure*}

Figure \ref{fig:nep_vs_power} presents the measured NEP as a function of incident power for our representative population of 490 resonators. The data neatly describe a transition between two NEP regimes, a thermal GR noise dominated regime where $\rm NEP^2 \approx K^2$ and a photon noise dominated regime where the shot and photon-generated quasi-particle recombination terms in Eq. \ref{eq:nep_inc} take over \cite{deVisser2014}. At expected flight loadings ($\sim 200 \ \mathrm{fW}$), detector noise contributes $\approx15\%$ of the total noise. The best fit parameters to Eq. \ref{eq:photon_nep_abs} give us an estimate of the median thermal GR noise floor of the detectors, $K=1.1\times10^{-17}$ $\mathrm{W/\sqrt{Hz}}$, and by ratiometrically comparing the median measured photon noise line with the calculated noise contribution from photons we can determine a median optical efficiency \cite{Janssen2013}, $\eta_\gamma = 0.67$ for our optical system. While the $\sim70$\% efficiency metric is a marked improvement over the $\sim 40$\% efficiency found for predecessor prototype in \j23.  The absorber is expected to have a 90\% band-averaged absorption efficiency \cite{Nie2022}. The additional losses might be attributed to roughness in the waveguide and/or misalignments between the multiflare angle horn and single-mode waveguide, which in combination can result in up to 30\% reduction of power transmission \cite{NiePhDthesis}.


\section{Discussion}
\label{sec:discussion}

The measurements presented here extend previous single-pixel demonstrations of the TIM long-wavelength detector architecture to a large-format, flight-representative array. The single-pixel measurements established that the chain-link aluminum LEKID design can achieve photon-noise limited performance at TIM-like loading. The present work demonstrates that this behavior persists for a statistically representative kilopixel-scale population when read out using the RFSoC-based system intended for the flight instrument.

The dark measurements show that the population-averaged thermal response follows the Mattis-Bardeen expectation for aluminum KIDs. The measured fractional frequency shifts agree well with the previous single-pixel measurements. The measured quality-factor evolution is weaker than in the earlier measurements. We interpret this
difference cautiously. It may reflect improved microwave drive-power optimization, differences in the optical and RF packaging, or the fact that the present measurement averages over a much larger population. Because the optical NEP is derived from the measured responsivity and measured noise, the sensitivity result does
not rely on the thermal-response model.

The optical measurements show that the detectors respond as expected to blackbody loading; for the purposes of flight-time readout, this sort of modest spread in $dx/dP$ is encouraging. This suggests that a per-pixel response curve can be characterized across flight-like loadings, then tracked and periodically calibrated in flight. This demonstrates the viability of our array-scale chip as a far-infrared detector. Additionally, our detector responsivities are well described by a two-parameter quasiparticle response model consisting of a thermal GR detector noise floor and a monotonic photon/Poisson contribution. At flight-like loading, nearly all of the characterized detectors show photon-limited scaling and satisfy the TIM long-wavelength sensitivity requirement. Many of the remaining outliers are strongly correlated with readout and analysis pathologies rather than with the intrinsic detector design, although a few ($<2\%$) detectors show severely degraded quality factor. 

Therefore the primary work remaining before flight is not a redesign of the LEKID absorber, but the development of a robust array-level measurement pipeline. Large-format KID arrays place a significant burden on the readout system: each tone must be placed accurately, driven near its loading-dependent optimal microwave power, and tracked as the resonance moves under changes in optical loading and stage temperature. In these data the dominant analysis failures are associated with
\begin{enumerate}
    \item resonators with near neighbors or unresolved collisions,
    \item resonators whose optimal drive power differs significantly from the
    population average (largely owing to degraded quality factor),
    \item resonators whose fitted circle parameters become unstable at some optical loadings.
\end{enumerate}

These effects motivate an automated characterization pipeline that jointly optimizes tone placement, drive power, collision identification, and resonator
tracking. Such a pipeline is essential for extending the present analysis from a conservative, statistically representative subset to the full flight focal plane. Substantial work toward this effort has already been undertaken. We present, methods for collision de-confusion in \cite{Liu2024}, and follow-up studies will present optimized resonator biasing procedures.

\section{Conclusion}
\label{sec:conclusion}

We have characterized a statistically representative population of 490 horn-coupled lumped-element kinetic inductance detectors from an 864-pixel flight-time science array for the long-wavelength module of the Terahertz Intensity Mapper. VNA sweeps recover approximately 759 resonators, or about 88\% of the array, while the conservative 490-detector analysis population excludes resonator collisions and poorly isolated frequency neighbors.

The detectors show both thermal and optical response consistent with Mattis-Bardeen theory and response models commonly used in literature. Using simultaneous RFSoC readout, we measure fractional-frequency noise and convert it to incident-power-referenced NEP using the measured detector responsivity. At flight-like optical loading, the majority of the characterized detectors are photon-noise dominated with a $\sim 15\%$ contribution  of the detector noise, which will be sufficient for TIM's key scientific objectives \cite{bracks26}.

These results demonstrate that the TIM long-wavelength LEKID architecture can be scaled from single-pixel and small-array demonstrations to kilopixel-class focal planes without losing the detector sensitivity needed for TIM's line-intensity mapping science. The main remaining limitations are associated with large-array readout operations: microwave drive-power optimization, robust tone tracking, and automated identification of frequency-domain collisions. Addressing these issues in an automated characterization pipeline will enable full-array sensitivity statistics for the TIM science arrays and provide a mature measurement framework for future far-infrared KID instruments.




\bibliographystyle{IEEEtran}
\bibliography{Kilopixel}

\end{document}